# Theoretical study of δ-5 boron monolayer as an anode material for Li and non-Li ion batteries

**Ajay Kumar[1] and Prakash Parida[1,a)]**

[1] *Department of Physics, Indian Institute of Technology Patna, Bihta, Patna, Bihar, India, 801106*
[a)] *Address all correspondence to these authors. e-mails: pparida@iitp.ac.in*

**Abstract**

We have studied the electrochemical performance of the δ-5 boron monolayer as an anode material for alkali metal (AM) and alkaline-earth metal (AEM) ion batteries using density functional theory simulations. The electronic properties, adsorption, diffusion rate, and storage behavior of various metal atoms (M) in the δ-5 boron monolayer are explored. Our study shows that the δ-5 boron monolayer possesses high electrical conductivity and a low activation barrier for electron and metal ion transit (0.493-1.117 eV), indicating a fast charge/discharge rate. Furthermore, the theoretical capacities of the δ-5 boron monolayer for Li, Na, and K are found to be greater than those of commercial graphite. The average open-circuit voltage for AM and AEM is reasonably low and in the range of 0.34-1.30 V. Our results show that δ-5 boron monolayer could be a promising anode material in lithium-ion and non-lithium ion rechargeable batteries.

## Introduction

Rechargeable lithium-ion batteries (LiBs) are essential components of a wide range of energy devices, from portable electronic gadgets to electric cars, and may soon provide superior eco-friendly transportation and good energy storage for renewable energy sources [1-3]. However, the earth's lithium supplies are limited: at the present consumption rate of 20 tons per year, the lithium sources can only be sustained for up to 65 years [4-5]. Further, the primary benefit of Na-ion batteries is the natural abundance and cheaper prices of sodium when compared to lithium. The abundance of Na to Li in the earth's crust is 23600 ppm to 20 ppm, and the cost of separation and purification of Na is less than that of Li [6-7]. But the significant disadvantage, other than Li-ion batteries, is the heavier mass of ions like Na, K, Mg, and Ca. Due to bulky ions, mobility will decrease, and the energy capacity of batteries will reduce because it is inversely proportional to molar mass. Many other alkali metals and alkali earth metals are being explored since they are more plentiful and have comparable electrochemical processes with LIBs [8-9]. Nonetheless, the superior performance of metal-ion batteries is restricted by a shortage of appropriate anode materials with cheap cost and large capacity, a quick charge-discharge time, and long cycle life, among other factors [10-11].

Two-dimensional (2D) materials may be one of the better choices for the anode in LiBs and non-lithium ion batteries (NLiBs). Their revolutionary chemical and physical characteristics and a high surface-to-volume ratio are promising as ideal anode materials for LiBs and NLiBs. Graphene, phosphorene, silicene, transition metal oxides (TMOs), chalcogenides (TMDs), and MXenes have been explored as anode materials [12-19]. While monoatomic layers are a more attractive candidate for anode materials, they are relatively chemically inert and did not affect much by external perturbation. The major disadvantage is the experimental synthesis limiting electrochemical performance [20-24].

We choose a boron-based monolayer known for its electrons deficient structure to overcome the limitation of graphitic material -a carbon-based anode material. The significant advantage of choosing the boron-based monolayers are 1) boron-based monolayers are electron-deficient and 2) the majority of boron allotropes are highly conducting due to their metallic nature [25-26]. In this work, we choose one of the boron allotrope (triangular motif boron monolayer), δ-5 boron monolayer. The δ-5 boron monolayer consists of six boron atoms per unit cell (rhombus), whose repetition along lattice vectors gives the hexagonal pore ring and a triangular motif. The δ-5 boron monolayer was first theoretically reported by Hui Tang in 2010 and was nomenclatured according to the hexagonal hole density ($\eta=1/7$)[27-28]. In 2012, Xiaojun wu also reported the electronic and structural stability of the same δ-5 boron monolayer with other boron structures[29]. Recently, the Dirac nature of electronic band structures in a few planar boron allotropes with honeycomb topology was reported by Wen-Cai Yi & others [30]. δ-5 boron monolayer is one of the boron allotropes reported for its linear response in the electronic bands spectrum. However, their electrochemical performances are yet to be investigated, which is critical for determining their application in LiBs and NLiBs.

We focus on the efficiency of the δ-5 boron monolayer in terms of electrochemical properties based on energy density and specific power to use it as an anode material in LIBs and NLiBs. Apart from voltage dependency, the energy density at electrochemical balance is related to storage capacity, whereas power density is related to metal atom diffusion over the δ-5 boron monolayer. To assess the electrochemical performances of the above-described systems, DFT investigations on the electronic characteristics of three extremely stable sites of δ-5 boron monolayer for different metal atoms are to be studied. Theoretical features of boron monolayers as an anode; storage capacity, open-circuit voltage and the diffusion barrier were also studied. Furthermore, the metal adsorbed monolayer durability is studied at a finite temperature up to 500K. We also estimate the metal atom concentration (x) per unit cell of the δ-5 boron monolayer and their corresponding storage capacity, open-circuit voltage and diffusion barriers for both lithium and non-lithium ion-based batteries such as sodium (Na), potassium (K), magnesium (Mg), and calcium (Ca).

## Results and discussion

### Structural stability and electronic properties of the δ-5 boron monolayer

The δ-5 boron monolayer has the P6/m space group. All the boron atoms are coordinated with the other five boron atoms. This structure follows the honeycomb topology; instead of simple hexagons in the graphene-like honeycomb structure, the δ-5 boron monolayer is composed of triangular and hexagonal motifs.

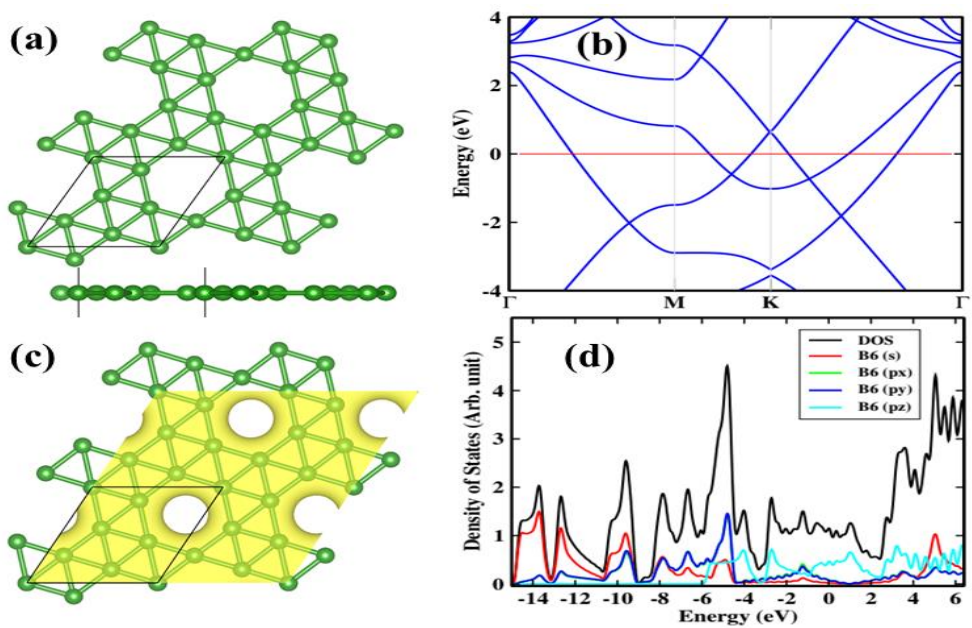

Figure 1: (a) The optimized structure (top and side views), (b) band structure, (c) charge density distribution, and (d) projected density of states (PDOS) of the δ-5 boron monolayer.

The Fermi level is scaled to zero.

Figure 1(a) shows the optimized geometry of the pristine δ-5 boron monolayer. Its cell parameters a=b=4.461Å and γ= 60º are comparable with previously reported theoretical values [27,29]. The primitive cell contains six boron atoms shown in figure 1(a). There is a uniform distribution of charges over the whole boron atoms belt with a hollow circular ring in a pristine δ-5 boron monolayer, as can be seen in figure 1(c). The calculated B-B bond lengths are found to be 1.68 Å throughout the monolayers, while this bond length changes depending upon the adsorption site of metal atoms. B-B bond length varies from 1.68 to 1.88 Å for the case of metal adsorbed δ-5 boron monolayer.

The mechanical stability of crystal structure has been described using stiffness constants. These elastic constants are derived from the first derivatives of stress to strain tensor. The δ-5 boron monolayer has a hexagonal P6/m space group. Consequently, it only has two independent elastic coefficients, $C_{11}$ and $C_{12}$. According to Born Criteria, $C_{11}$, $C_{12}$ and $C_{11} - C_{12} > 0$ for a 2D hexagonal structure. For the δ-5 boron monolayer, the computed values of $C_{11}$, $C_{12}$, Y(Young's modulus), and υ (Poisson's ratios) are found to be 108.52, 20.32, 104.71, and 0.18 GPa, respectively. δ-5 boron monolayer is less stiffer than both graphene and single-layer h-BN but mechanically as stable as MoS2[14,19].

To check the thermal stability of the pristine δ-5 boron monolayer, we have considered a 2×2×1 supercell. We have performed ab initio molecular dynamics (AIMD) simulations on δ-5 boron monolayer at different temperatures of 100 K, 300 K and 500 K using the VASP Package. The structural stability and planarity of δ-5 boron monolayer is retained in a 5 ps molecular dynamic simulation as shown in figure S1(a) and S2(a). We also have computed the phonon dispersion spectrum (shown in figure S3) to verify the dynamic stability of the pure δ-5 boron monolayer. Absence of any negative frequency in our calculation suggests the dynamical stability of δ-5 boron monolayer.

The electronic properties of electrode materials are strongly connected to battery cyclability and rate performance. The electronic band spectrum of the pristine δ-5 boron monolayer shows metallic nature as a few number of bands cross the Fermi level, which can be seen in figure 1(b). The s-orbitals of B atoms contribute a little to the electronic states near the Fermi level, while the p-orbitals are primarily responsible for its metallic nature. Furthermore, there are a few Dirac points where the bands intersect linearly. Two of those are at 3.46 eV, 1.162 eV below the Fermi level, and the third one is at 2.94 eV above the Fermi level. The total density of state (DOS) in figure 1(d) shows non-vanishing states near the Fermi level, indicating the high electron conductivity in the δ-5 boron monolayer. The metallic nature of the δ-5 boron monolayer provides an inherent advantage for electrical conductivity and satisfying electrochemical properties for improved battery cycling.

**Adsorption of single metal atom (M, M=Li, Na, K, Mg and Ca) on the δ-5 boron monolayer**

To study the storage capacity and diffusion of metal atoms on the surface of the δ-5 boron monolayer, we have considered four possible adsorption sites of metal atoms M atoms (M = Li, Na, K, Mg, and Ca): hexagon-site (H-site), Trigonal-site (T-site) Boron-site (B1-site), and bridges-site (B2-site) as shown in figure 2(a).

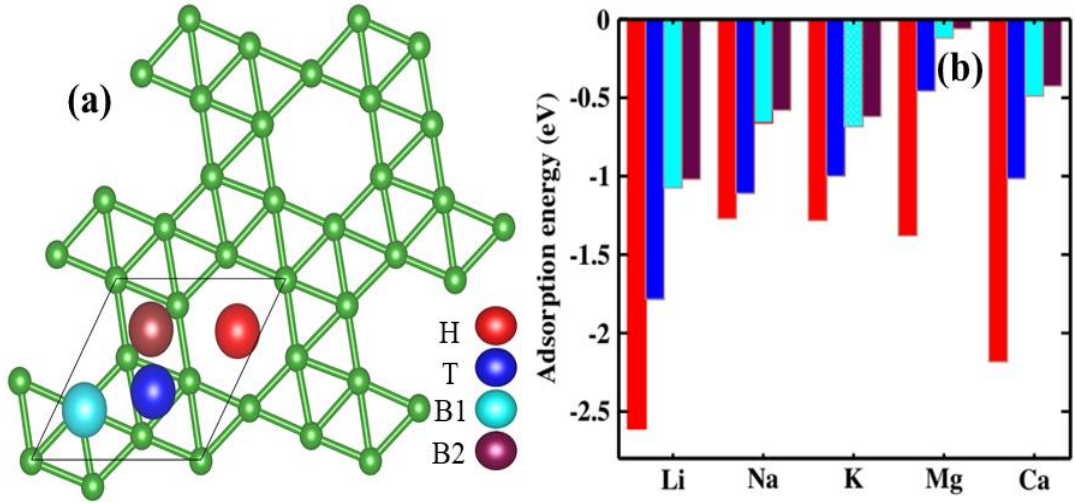

Figure 2: (a) Adsorption sites H, T, B1, and B2 are denoted in red, blue, cyan and maroon colour ball, respectively. (b) Corresponding adsorption energies for a single metal atom (M = Li, Na, K, Mg, and Ca) on the δ-5 boron monolayer.

For stability analysis, we have calculated adsorption energy using the following formula:

$$E_{ads} = E_{M/\delta-5} - E_{\delta-5} - E_M \qquad (1)$$

Where $E_{M/\delta-5}$ and $E_{\delta-5}$ are the energies of the metal adsorbed δ-5 boron monolayer and pristine δ-5 boron monolayer, respectively, and $E_M$ represents the energy of an isolated M atom. Because it follows an exothermic process, the negative adsorption energy indicates that the metal adsorbed boron sheets are stable. The negative $E_{ads}$ values in figure 2(b) demonstrate that almost all metal atoms can be efficiently adsorbed on δ-5 boron monolayer. Site H is the most energetically favourable adsorption site for all metal atoms, implying that it is the most symmetric site for metal to feel relaxed compared to other sites. The equivalent $E_{ads}$ values at site H for Li Na, K, Mg, and Ca are -2.619, 1.717, -1.382, -0.831, and -2.184 eV, respectively. Negative adsorption energies ensure that metal atoms adsorb strongly on the δ-5 boron monolayer. Adsorption of metal atom at B2-site (bridge site) is the most unstable.

Further, we used a 2×2×1 supercell to avoid the interaction between two adjacent metal atoms. The distance between two metal atoms is 9.416 Å when calculating the adsorption energy of a metal atom on a boron monolayer. The interaction energy, $E_{\text{int}}$, between adjacent metal atoms is defined as [31]

$$E_{\text{int}} = E_{supercell(n \times n \times 1)} - E_{isolated\ atom}$$

Where $E_{supercell(n \times n \times 1)}$ is the energy of the metal atom calculated by placing the metal atom with the same distance ($R_{M-M}$=9.416 Å) as maintained to calculate the adsorption energy and $E_{isolated\ atom}$ is the energy an of isolated metal atom calculated by placing a metal atom at the centre of a cube with the dimensions of lattice constant of 15 Å. The interaction energy while considering 2×2×1 supercell for all metal atoms is negligible compared to their adsorption energy reported in table S1. It implies that, 2×2×1 supercell is essential and sufficient to avoid interaction between adjacent metal atoms.

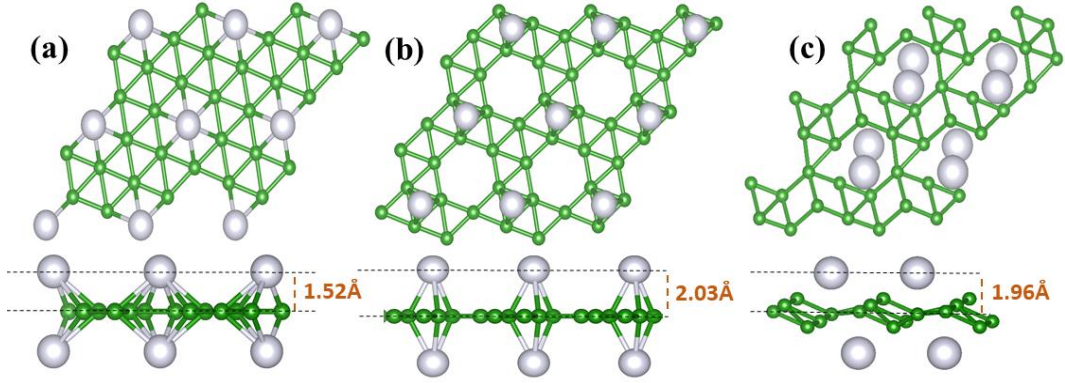

Figure 3: The optimized structures for the Li-adsorbed boron monolayer at sites (a) H, (b) T, and (c) B1 are shown.

The adsorption height in a vertical direction is the distance between a boron layer and a metal atom at different sites. For Li, the adsorption height at the H, T and B sites are 1.48, 1.86 and 2.05 Å as shown in figure 3. Furthermore, because of different atomic radii, the adsorption height varies with different atoms, even if they belong to the same group. For the most stable site H, lithium has the lowest $E_{ads}$ value.

To investigate the charge transfer between the metal atom and δ-5 boron monolayer, we have calculated the charge density difference between pristine and adsorbed δ-5 boron monolayer using the following formula.

$$\rho_{net} = \rho_{(M/\delta-5)} - \rho_{\delta-5} - \rho_M \quad (2)$$

Where $\rho_{(M/\delta-5)}$ and $\rho_{\delta-5}$ denote the electron densities of the relaxed δ-5 boron monolayer with absorbed metal atom and pristine δ-5 boron monolayer, respectively, and $\rho_M$ is the electron density of the metal atom. To visualize it, we have plotted charge density difference distribution in figure 4(a) for lithium and in 4(b) for Mg adsorbed at site H. The plot shows that, all adsorbed metal atoms act as electron donors, transferring electrons to the δ-5 boron monolayer shown in figure S4. The yellow color region between the metal atom and monolayer confirmed the charge polarization towards the boron monolayer. The cyan color on metal atoms implies that, a fraction of charge transfer occurs from metal atoms to the δ-5 boron monolayer.

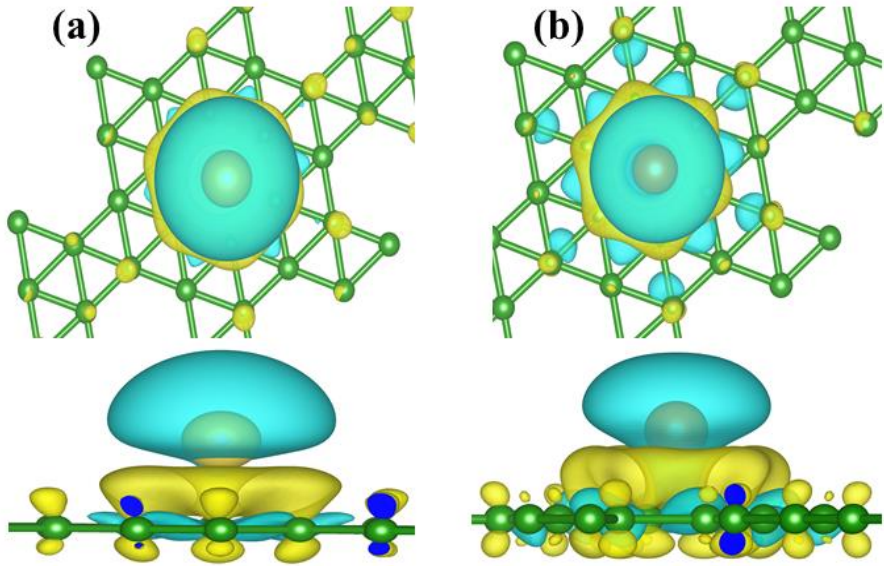

Figure 4: Charge density difference for (a) Li and (b) Mg adsorbed on δ-5 boron monolayer at site H with iso-surface value of 0.002 eÅ$^{-3}$.

According to Bader's charge analysis, the charge transfer from Li, Na, K, Mg, and Ca to the δ-5 boron monolayer at site H are 0.113, 0.316, 0.708, 1.153, and 1.189 electronic charge. The charge transfer is very low for the lithium because of the minor electronegativity difference between Li and B.

We have investigated the electronic band structure characteristics to check the conductivity of the δ-5 boron monolayer with metal atom adsorption. Upon adsorption of metal atoms, there is hardly any change to the overall nature of the band structure of the pristine δ-5 boron monolayer. Because of charge transfer from metal atoms to boron monolayer, only an up-shift in Fermi level is noticed, as shown in figure S5. The overall metallic nature of the boron monolayer is retained with the adsorption of metal atoms.

To check the thermal stability of the metal adsorbed δ-5 boron monolayer, we have considered a 2×2×1 supercell. We have performed ab initio molecular dynamics (AIMD) simulations on metal absorbed δ-5 boron monolayer up to four layers at different temperatures of 100 K, 300 K and 500 K with 5ps using the VASP package. Figure S1 shows that, for adsorption of all metal atoms, the planarity of the boron monolayer is not destroyed and structure is quite stable for temperature up to 300K. However, at 500 K, there is a little distortion in planarity of boron monolayers, as shown in figure S2.

**Diffusion of metal atoms on the δ-5 boron monolayer**

The charging/discharging rate substantially correlates with the transport characteristics of the ions and electrons, depending on the diffusion of metal atoms and the electrical conductivity of the δ-5 boron monolayer. We therefore have studied diffusion process of metal atoms on the surface of the δ-5 boron monolayer. The underlying structural symmetry of the boron monolayer helps to determine the diffusion pathway. As the H site is the most stable adsorption site, we have considered the diffusion only from H to H site via the one bridge site (or two nearby trigonal sites). In the path, the H-T-B2-T-H metal atom moves across two T sites and one bridge site, as shown in figure 5(a) for the Li atom. All other metal atoms follow a similar diffusion route. According to our estimation, this path has substantial diffusion barriers mb(0.493–1.785 eV) for all metal atoms. The δ-5 boron monolayer has diffusion barriers of 0.493, 0.585, and 1.117 eV for K, Na, and Li atoms, respectively. It is noted that higher activation energy occurs due to the strong electron repulsion at the B2-site. Changyan Zhu et al. reported the diffusion height in the range of 0.53-1.23 eV for Li atom in $B_7P_2$ monolayer. They represented three different paths out of which the Li atom faced the highest barrier through the bridge site and the least through the phosphorus site [32]. Similarly, for Graphene, Li faces a barrier of 0.324 eV [33], which is approximately the difference between adsorption energy at the stable Hexagon centre and the unfavourable bridge site. Similarly, in our case, the barrier height is the highest at bridge position, B2-site. As the adsorption energy difference between H-site and B2-site is the largest for Li atom, the barrier height is the largest for Li compared to other atoms. This is because, compared to other atoms, Li atom adsorps strongly at the H-site. Hence, diffusion of Li atom through any other site except H-site has to face a huge barrier. However, as reported, the δ-5 boron monolayer has enough ion mobility to access the above diffusion barriers for Li, Na, and K [34]. Compared to the monovalent alkali metal atoms (Li, Na, and K), the diffusion barriers for Mg and Ca are 1.484 and 1.785 eV, respectively, which are substantially higher. We also compare the impact of vdW interaction on the diffusion barrier in table S2. It has been noticed that vdW interaction lowers the total energy of the system with a minor change in the diffusion barrier for all metal atoms except Li 1.171 (1.72) and Ca 1.785 (1.572, respectively.

**Open-circuit voltage and theoretical storage capacity**

The charge-discharge process calculates the average open-circuit voltage (OCV) and storage capacity (C) of rechargeable metal-ion batteries. The following half-cell reaction may be used to illustrate the charge-discharge mechanism of a boron-based metal-ion battery:

$$M_x\delta - 5 \text{ boron monolayer} \leftrightarrow \delta - 5 \text{ boron monolayer} + xM^{n+} + xne^- \quad (3)$$

The average open circuit voltage is calculated using the Gibbs free energy change of this half-cell process (OCV).

$$\Delta G = \Delta E + P\Delta V - T\Delta S \quad (4)$$

As we deal with constant pressure and temperature, $P\Delta V$ is of the order of $10^{-05}$ eV and $T\Delta S$ term is negligibly small (thermal energy at room temperature) as compared to $\Delta E$. Generally, $\Delta E$ is of the order of 1-5 eV which is much higher than $T\Delta S$ contribution, which is of a few meV at room temperature [33]. Hence, the average OCV is calculated from the average adsorption energy ($E_{ave}$).

$$OCV = -E_{ave}/xne \quad (5a)$$

$$E_{ave} = E_{M_x\delta-5} - E_{\delta-5} - x E_M \quad (5b)$$

Where $E_{M_x\delta-5}$, $E_{\delta-5}$, and $E_M$ are the energies of the M-adsorbed δ-5 boron monolayer, δ-5 boron monolayer, metal atom M, respectively, and n denotes the number of electrons which is completely ionized from metal atoms. For alkali metals (Li, Na, K), n =1 and n=2 for alkali earth metals (Mg, Ca). x is the chemical composition of the M atom per unit cell and e is the electronic charge. We have considered different compositions with x=1, 2 and 4 to examine the adsorption capacities. For all adsorbed metal atoms, δ-5 boron monolayer retains the original structure up to concentration x= 4 except the Ca atom. The honeycomb topology of the δ-5 boron monolayer is distorted for Ca atom, as shown in figure 6 (b).

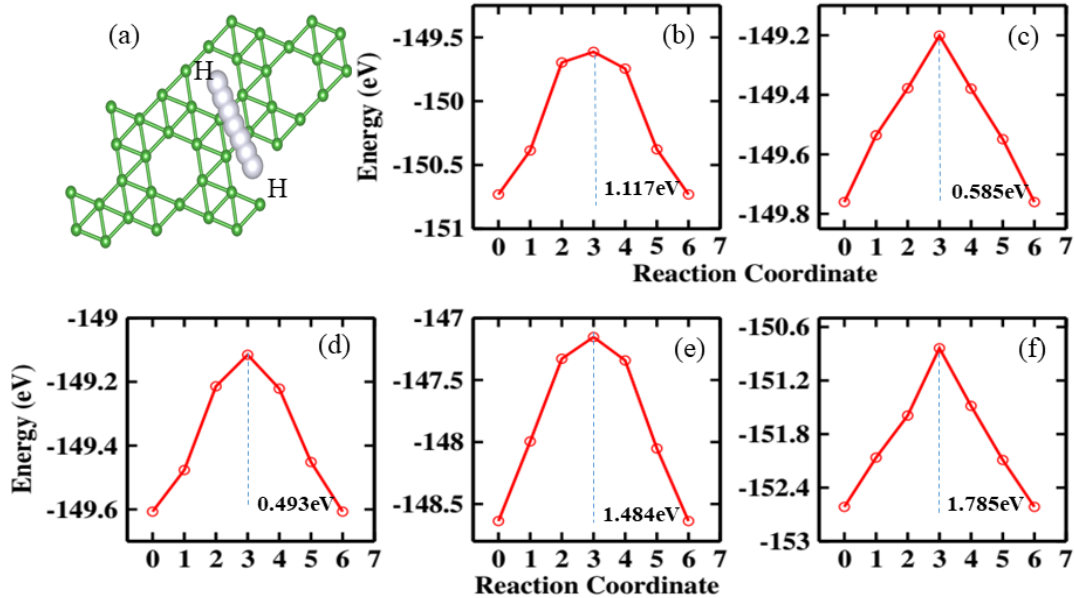

Figure 5: (a) The diffusion pathway for a metal atom on the δ-5 boron monolayer from one H site to another H site. The diffusion barrier for (b) Li, (c) Na, (d) K, (e) Mg, and (f) Ca.

The following equation [33] is used to compute the relevant adsorption capacities:

$$C = xnF/M_{Li_x\delta-5} \quad (6)$$

where C is storage capacity, x is the concentration of Metal atom per unit cell of δ-5 boron monolayer, n is the number of valence states of a metal atom, F is Faraday's constant and $M_{Lix\delta-5}$ is the molecular mass of metal adsorbed boron monolayer.

Further, on increasing the concentration of metal atom, it is noted that the height of metal atoms from boron monolayers follow the same trend that we observed in single-site adsorption. In case of lithium, for x=1, Li sit at H site, for x=2, there are two Li atoms layers below and above the boron sheet at H site and for x=4, There are four layers of Li atoms, with the last two Li atoms occupying the trigonal position. As a result, there are five atomic levels along the z-axis. For maximum x=4. Figure 6 shows a δ-5 boron monolayer intercalated with four Li atom layers, two above and two below the δ-5 boron monolayer.

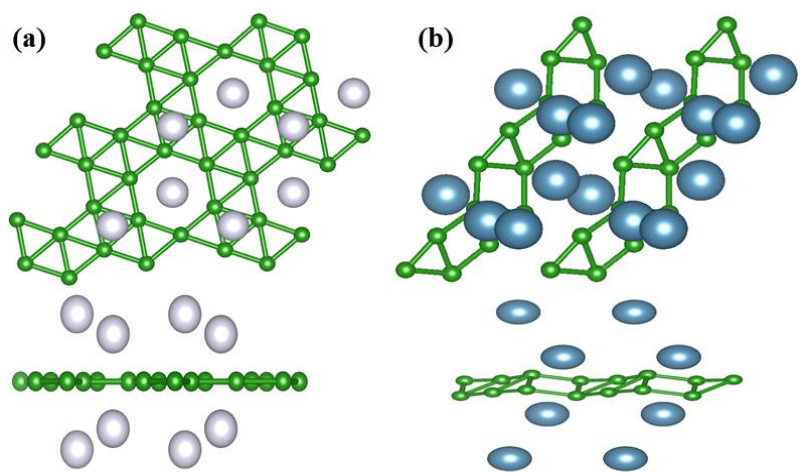

Figure 6: Top and side view of optimized structures for (a) Li$_4$B6 and (b) Ca$_4$B6.

For various x values, we have analyzed the stability of adsorption of all metal atoms on the δ-5 boron monolayer. Our calculations reveal that the varying content of metal atoms goes up to 4 for alkali metal and Mg metal. Still, in the case of Ca, increasing the x, δ-5 boron monolayer has been distorted and failed to maintain for x= 4 shown in figure 6. Therefore, For Ca, only x=1 and 2 electrochemical investigations are suitable. The adsorption energy equation (1) is further used to investigate the stability of Li$_4$B6, Na$_4$B6, K$_4$B6, Mg$_4$B6, and Ca$_2$B6. The lower thermodynamic stability of M$_x$B6 is attributable to electrostatic repulsion among the adsorbed metal atoms. The E$_{ave}$ values for Li, Na, K and Mg are still more negative, as shown in the table, which favour stacking the metal atoms up to two layers across both surfaces of the δ-5 boron monolayer.

Further, we also reported the layer-by-layer average adsorption energies in table 1. In general, on increasing the metal atom layers, E$_{ave}$ should decrease to avoid the creation of metal clusters[35]. We also get a similar trend in our calculation. For all metal atoms except Ca, up to four layers, there is a decrease in the adsorption energy. Adsorption energy for 3-layer and 4-layer E$_{ave}$ energy increases only for Ca atom. A similar dispensary is shown in figure S6, where Ca atoms layers break the topology of the δ-5 boron monolayer. Hence, the highest value of x for Li, Na, K and Mg is 4 while for Ca, x is 2.

| Metal atom | $E_{ave/single}$ (eV) | $E_{ave/layer-2}$ (eV) | $E_{ave/layer-4}$ (eV) |
|---|---|---|---|
| Li | -2.619 | -0.265 | -0.0191 |
| Na | -1.717 | -0.057 | -0.0185 |
| K | -1.382 | -0.163 | -0.125 |
| Mg | -0.831 | -0.114 | -0.089 |
| Ca | -2.184 | -0.421 | - |

Table 1: - The calculated layer-by-layer average adsorption energy $E_{layer-n}$ for metal adsorbed δ-5 boron monolayer.

Finally, using equations (5a), (5b), and (6), we have estimated the average open-circuit voltage (OCV) and storage capacity for M-ion batteries at varied x values. The estimated OCVs are presented in figure 7 as a function of adatoms Li, Na, K, Mg, and Ca concentration (x) in $M_xB6$. The storage capacity on the anode for all M-ion batteries rises as x value increases, but the average OCV on the anode decreases. The maximal specific capacities of Li, Na, K, Mg, and Ca are determined to be 1157.35, 683.44, 484.52, 661.413, and 476.09 mAhg$^{-1}$, respectively. Compared to other 2D anode materials, such as graphite[36-37] (372 mAhg$^{-1}$ for Li, 284 mAhg$^{-1}$ for Na, and 273mAhg$^{-1}$ for K), phosphorene [19] (389.02 mAhg$^{-1}$, 315.52 mAhg$^{-1}$, and 310.71 mAhg$^{-1}$ for Li, Na, and K), and MXenes (447.8 mAhg$^{-1}$ for Li, 351.8 mAhg$^{-1}$ for Na, 191.8 mAhg$^{-1}$ for K)[38-39]. These values are found to be higher for δ-5 boron monolayer. We have calculated average OCVs for Li, Na, and K by utilizing the dispersion interaction (vdW interaction) to avoid the dynamical fluctuating charge distribution. An informative comparative with and without considering vdW interaction is reported in table S2.

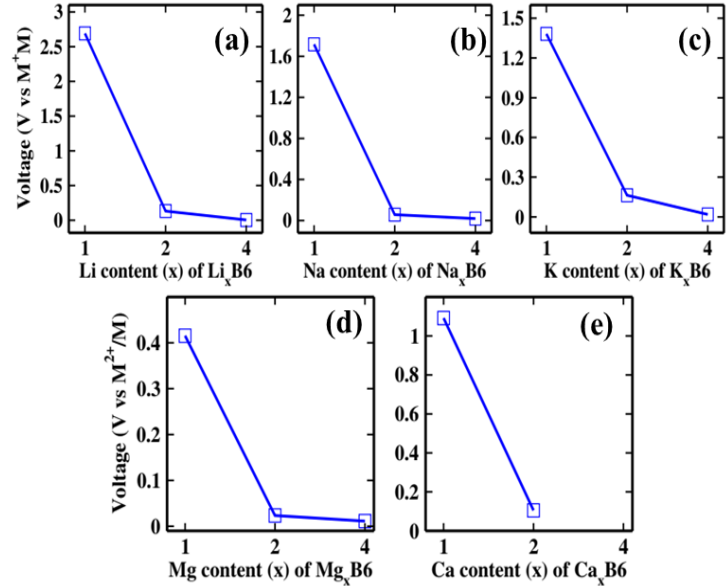

Figure 7: Voltage profiles of δ-5 boron monolayer for metal atoms by varying the concentration (x).

The average calculated OCVs are reasonable and within a moderate range (0.96, 0.59, 0.55, 0.34, and 1.302 V for Li, Na, K, Mg and Ca respectively). A mild value of OCV for anode is practically good since it lowers the chance of dendritic formation [40-41]. The typical OCV for Li is between that of commercial anode materials, ranging from 0.11 V for graphite [42-43] to 1.50–1.80 V for TiO$_2$ [44]. The average voltage of Na is likewise within the 0.00–1.00 V acceptable voltage range of the sodium anode [45]. In comparison, the usual voltage of K is roughly 0.17 V, which is closer to the voltage of K insertion in graphite at roughly 0.17 V [46].

The computed OCVs for Li, Na, and K on the δ-5 boron monolayer are smaller than those in other reported boron sheets (1.0V) [47]. Due to these small OCV values, δ-5 boron monolayer might be used as an anode for LiBs, NaBs, and KBs. Based on the computed parameters of average open-circuit voltage and specific capacity, the δ-5 boron monolayer could be a good anode material for Li, Na, and K-ion batteries.

## Conclusions

We have investigated the possibility of employing an inherently metallic boron porous network as an anode material for metal (Li, Na, K, Mg, and Ca) ion batteries using DFT simulations. Our results show that the δ-5 boron monolayer is best suited as an anode material for Alkali-ion batteries because it can adsorb up to four (two above and two below) layers of Li, Na and K. The calculated maximum specific capacities are found to be 1157.35, 683.44, and 484.52 mAhg$^{-1}$ for Li, Na, and K, respectively. The activation energies for Li (1.117 eV), Na (0.585 eV), and K (0.493 eV) atoms show fast diffusion on the δ-5 boron monolayer. Although, for Li ion, δ-5 boron monolayer has a high diffusion barrier, its high storage capacity will have an advantage over other anode material. Furthermore, the average OCVs of the δ-5 boron monolayer for LiBs, NaBs, and KBs are relatively low, falling within a reasonable range (0.344-1.30 V). We conclude that the δ-5 boron monolayer has tremendous promise as a high-performance anode material for alkali metal batteries due to its huge charge capacity, low energy barriers, excellent structural stability, and moderate OCV values.

## Computational methods

DFT-based computational techniques inside the Vienna ab initio simulation packages (VASP) are used for all the calculations. It employs the Projector Augmented Wave (PAW) approach for valence electron interactions with core electrons and periodic boundary conditions. The appropriate pseudopotentials are generated for B, Li, Na, K, Mg, and Ca atoms with valence electron electronic configurations $2s^2 2p^1$, [He] $2s^1$, [Ne] $3s^1$, [Ar] $4s^1$, [Ne] $3s^2$, and [Ar] $4s^2$, respectively. A plane wave basis with an energy cut-off of 600 eV and generalized gradient approximation (GGA) in the Perdew–Burke–Ernzerhof (PBE) exchange-correlation functional are used. Pristine boron monolayers and adsorbed metal boron monolayers are fully relaxed using a conjugate–gradient algorithm with force tolerance of 0.01 eV/Å on each atom. The van der Waals (vdW) interactions were also used in Grimme zero damping DFT-D3 method. To eliminate interactions between the periodic layers, a vacuum space greater than 15Å is used along the Z-axis. A Monkhorst–Pack scheme of 25×25×1 k-points of the 2×2×1 supercell is employed in all our calculations. We use Xcrysden to determine the high symmetry path (Γ-M-K-Γ) in the first brillouin zone. The transition state (TS) structure and diffusion barrier of metal adsorbed δ-5 boron monolayer are calculated using climbing-nudged elastic band (C-NEB) techniques. We investigate mechanical properties using the strain-stress method (SSM) using VASP. The dynamics stability of the B6-monolayer is studied by using a Phonopy interface with VASP. We use the finite displacement method by creating a $3 \times 3 \times 1$. supercell Thermal stability of pristine and metal adsorbed boron monolayer are investigated using ab initio molecular dynamics (AIMD) simulations within the NVT ensemble at 100K, 300K, and 500K. We have utilized the VESTA software to show the atomic structures and charge densities.

## Acknowledgement

AK thanks University Grants Commission (UGC), New Delhi, for financial support in the form of a Junior Research Fellowship (DEC18-512569-ACTIVE). PP thanks DST-SERB for ECRA project (ECR/2017/003305).

## Data availability

The datasets generated during and/or analyzed during the current study are available from the corresponding author on reasonable request.

## Code availability

Not applicable

## Declarations

**Conflict of interest** On behalf of all authors, the corresponding author states that there is no conflict of interest.

## References

1. J.P. Barton and D.G. Infield: Energy storage and its use with intermittent renewable energy *IEEE transactions on energy conversion.* **19**(2), 441 (2004).
2. M. Lowe, S. Tokuoka, T. Trigg and G. Gereffi: Lithium-ion batteries for electric vehicles *The US Value Chain, Contributing CGGC researcher: Ansam Abayechi.* (2010).
3. N. Nitta, F. Wu, J.T. Lee and G. Yushin: Li-ion battery materials: present and future *Materials today.* **18**(5), 252 (2015).
4. A. Ponrouch and M.R. Palacín: Post-Li batteries: promises and challenges *Philosophical Transactions of the Royal Society A.* **377**(2152), 20180297 (2019).
5. C. Delmas: Sodium and sodium-ion batteries: 50 years of research *Advanced Energy Materials.* **8**(17), 1703137 (2018).
6. M.D. Slater, D. Kim, E. Lee and C.S. Johnson: Sodium-ion batteries *Advanced Functional Materials.* **23**(8), 947 (2013).
7. Y. Wu and Y. Yu: 2D material as anode for sodium ion batteries: Recent progress and perspectives *Energy Storage Materials.* **16**, 323 (2019).
8. Y. Liu, B.V. Merinov and W.A. Goddard: Origin of low sodium capacity in graphite and generally weak substrate binding of Na and Mg among alkali and alkaline earth metals *Proceedings of the National Academy of Sciences.* **113**(14), 3735 (2016).
9. B. Ji, F. Zhang, X. Song and Y. Tang: A novel potassium-ion-based dual-ion battery *Advanced materials.* **29**(19), 1700519 (2017).
10. K. Chayambuka, G. Mulder, D.L. Danilov and P.H. Notten: Sodium-ion battery materials and electrochemical properties reviewed *Advanced Energy Materials.* **8**(16), 1800079 (2018).
11. E. Peled: The electrochemical behavior of alkali and alkaline earth metals in nonaqueous battery systems—the solid electrolyte interphase model *Journal of The Electrochemical Society.* **126**(12), 2047 (1979).
12. J.K.S. Goodman and P.A. Kohl: Effect of alkali and alkaline earth metal salts on suppression of lithium dendrites *Journal of The Electrochemical Society.* **161**(9), D418 (2014).
13. L. Shi and T. Zhao: Recent advances in inorganic 2D materials and their applications in lithium and sodium batteries *Journal of Materials Chemistry A.* **5**(8), 3735 (2017).

14. L. Wang, Z. Wei, M. Mao, H. Wang, Y. Li and J. Ma: Metal oxide/graphene composite anode materials for sodium-ion batteries *Energy Storage Materials.* **16**, 434 (2019).
15. M. Gu, Y. He, J. Zheng and C. Wang: Nanoscale silicon as anode for Li-ion batteries: The fundamentals, promises, and challenges *Nano Energy.* **17**, 366 (2015).
16. C. Zhang, M. Yu, G. Anderson, R.R. Dharmasena and G. Sumanasekera: The prospects of phosphorene as an anode material for high-performance lithium-ion batteries: a fundamental study *Nanotechnology.* **28**(7), 075401 (2017).
17. H. Hwang, H. Kim and J. Cho: MoS2 nanoplates consisting of disordered graphene-like layers for high rate lithium battery anode materials *Nano letters.* **11**(11), 4826 (2011).
18. Y. Jiang, M. Hu, D. Zhang, T. Yuan, W. Sun, B. Xu and M. Yan: Transition metal oxides for high performance sodium ion battery anodes *Nano Energy.* **5**, 60 (2014).
19. Q. Tang, Z. Zhou and P. Shen: Are MXenes promising anode materials for Li ion batteries? Computational studies on electronic properties and Li storage capability of Ti3C2 and Ti3C2X2 (X= F, OH) monolayer *Journal of the American Chemical Society.* **134**(40), 16909 (2012).
20. D. Chen, G. Ji, B. Ding, Y. Ma, B. Qu, W. Chen and J.Y. Lee: Double transition-metal chalcogenide as a high-performance lithium-ion battery anode material *Industrial & Engineering Chemistry Research.* **53**(46), 17901 (2014).
21. M. Aydinol, A. Kohan, G. Ceder, K. Cho and J. Joannopoulos: Ab initio study of lithium intercalation in metal oxides and metal dichalcogenides *Physical Review B.* **56**(3), 1354 (1997).
22. C.M. Julien, A. Mauger, K. Zaghib and H. Groult: Comparative issues of cathode materials for Li-ion batteries *Inorganics.* **2**(1), 132 (2014).
23. X. Liu, X. Zhu and D. Pan: Solutions for the problems of silicon–carbon anode materials for lithium-ion batteries *Royal Society open science.* **5**(6), 172370 (2018).
24. X. Chen and Y. Zhang: The main problems and solutions in practical application of anode materials for sodium ion batteries and the latest research progress *International Journal of Energy Research.* **45**(7), 9753 (2021).
25. L. Kong, K. Wu and L. Chen: Recent progress on borophene: Growth and structures *Frontiers of Physics.* **13**(3), 1 (2018).
26. L. Adamska, S. Sadasivam, J.J. Foley IV, P. Darancet and S. Sharifzadeh: First-principles investigation of borophene as a monolayer transparent conductor *The Journal of Physical Chemistry C.* **122**(7), 4037 (2018).
27. D. Liu and D. Tománek: Effect of net charge on the relative stability of 2D boron allotropes *Nano Letters.* **19**(2), 1359 (2019).
28. H. Tang and S. Ismail-Beigi: First-principles study of boron sheets and nanotubes *Physical Review B.* **82**(11), 115412 (2010).
29. X. Wu, J. Dai, Y. Zhao, Z. Zhuo, J. Yang and X.C. Zeng: Two-dimensional boron monolayer sheets *ACS nano.* **6**(8), 7443 (2012).
30. W.-c. Yi, W. Liu, J. Botana, L. Zhao, Z. Liu, J.-y. Liu and M.-s. Miao: Honeycomb boron allotropes with Dirac cones: A true analogue to graphene *The Journal of Physical Chemistry Letters.* **8**(12), 2647 (2017).
31. Zhang S, Hu X, Lu Q, Zhang J: Density Functional Theory Study of Arsenic and Selenium Adsorption on the CaO (001) Surface *Energy & Fuels.* **25**(7), 2932 (2011).
32. Zhu C, Lin S, Zhang M, Li Q, Su Z, Chen Z: Ultrahigh capacity 2D anode materials for lithium/sodium-ion batteries: an entirely planar B7P2 monolayer with suitable pore size and distribution. *Journal of Materials Chemistry A.* **8**(20), 10301 (2020).
33. Chan KT, Neaton JB, Cohen ML: First-principles study of metal adatom adsorption on graphene *Physical Review B.* **77**(23), 235430 (2008).
34. Adamska L, Sadasivam S, Foley IV JJ, Darancet P, Sharifzadeh S: First-principles investigation of borophene as a monolayer transparent conductor *The Journal of Physical Chemistry C.* **122**(7) 4037 (2018).

35. Yang Z, Zheng Y, Li W, Zhang J: Tuning the electrochemical performance of Ti3C2 and Hf3C2 monolayer by functional groups for metal-ion battery applications *Nanoscale*. **13**(26), 11534 (2021).
36. H. Xu, H. Chen and C. Gao: Advanced graphene materials for sodium/potassium/aluminum-ion batteries *ACS Materials Letters.* **3**(8), 1221 (2021).
37. Y. Zhang, X. Xia, B. Liu, S. Deng, D. Xie, Q. Liu, Y. Wang, J. Wu, X. Wang and J. Tu: Multiscale graphene-based materials for applications in sodium ion batteries *Advanced Energy Materials.* **9**(8), 1803342 (2019).
38. Y. Jing, Z. Zhou, C.R. Cabrera and Z. Chen: Metallic VS2 monolayer: a promising 2D anode material for lithium ion batteries *The Journal of Physical Chemistry C.* **117**(48), 25409 (2013).
39. Y. Jing, E.O. Ortiz-Quiles, C.R. Cabrera, Z. Chen and Z. Zhou: Layer-by-layer hybrids of MoS2 and reduced graphene oxide for lithium ion batteries *Electrochimica Acta.* **147**, 392 (2014).
40. Liu K, Zhang B, Chen X, Huang Y, Zhang P, Zhou D: Modulating the Open-Circuit Voltage of Two-Dimensional MoB MBene Electrode via Specific Surface Chemistry for Na/K Ion Batteries: A First-Principles Study *The Journal of Physical Chemistry C*. **125**(33), 18098 (2021).
41. Lv X, Li F, Gong J, Gu J, Lin S, Chen Z: Metallic FeSe monolayer as an anode material for Li and non-Li ion batteries: a DFT study *Physical Chemistry Chemical Physics*. **22**(16), 8902 (2020).
42. R. Yazami and Y.F. Reynier: Mechanism of self-discharge in graphite–lithium anode *Electrochimica Acta.* **47**(8), 1217 (2002).
43. K. Persson, V.A. Sethuraman, L.J. Hardwick, Y. Hinuma, Y.S. Meng, A. Van Der Ven, V. Srinivasan, R. Kostecki and G. Ceder: Lithium diffusion in graphitic carbon *The journal of physical chemistry letters.* **1**(8), 1176 (2010).
44. Y. Ren, Z. Liu, F. Pourpoint, A.R. Armstrong, C.P. Grey and P.G. Bruce: Nanoparticulate TiO2 (B): an anode for lithium-ion batteries *Angewandte Chemie.* **124**(9), 2206 (2012).
45. Mortazavi M, Wang C, Deng J, Shenoy VB, Medhekar NV: Ab initio characterization of layered MoS2 as anode for sodium-ion batteries *Journal of Power Sources*. **268**, 279 (2014).
46. Jian Z, Luo W, Ji X: Carbon Electrodes for K-Ion Batteries *Journal of the American Chemical Society*. **137**(36), 11566 (2015).
47. H. Jiang, Z. Lu, M. Wu, F. Ciucci and T. Zhao: Borophene: a promising anode material offering high specific capacity and high rate capability for lithium-ion batteries *Nano Energy.* **23**, 97 (2016).